# Dynamic Management Techniques
# For Increasing Energy Efficiency within a Data Center

**Iulia DUMITRU \*, Grigore STAMATESCU \***
**Ioana FĂGĂRĂȘAN \*, Sergiu Stelian ILIESCU \***
*\*Faculty of Automatic Control and Computers, POLITEHNICA University of Bucharest*
*E-mail: iulia.dumitru@ aii.pub.ro*
*grigore.stamatesc@ac.pub.ro*

**Abstract:** In ours days data centers provide the global community an indispensable service: nearly unlimited access to almost any kind of information we can imagine by supporting most Internet services such as: Web hosting and E-commerce services. Because of their capacity and their work, data centers have various impacts on the environment, but those related with the electricity use are by far the most important. In this paper, we present several power and energy management techniques for data centers and we will focus our attention on techniques that are explicitly tailored to servers and their workloads.

*Keywords:* data center; energy efficiency; energy management; virtualization; server's workload

1. INTRODUCTION

In ours days data centers provide the global community an indispensable service: nearly unlimited access to almost any kind of information we can imagine by supporting most Internet services such as: Web hosting and E-commerce services. Power and energy consumption are key concerns of Internet Data Centers. These centers house hundreds, sometimes thousands, of servers and supporting cooling infrastructures. Research on power and energy management for servers can ease data center installation, reduce costs, and protect the environment. Given these benefits, researchers have made important strides in conserving energy in servers. Inspired by this initial progress (Ricardo Bianchiniy and all), researchers are delving deeper into this topic.

The first step in prioritizing energy saving opportunities is to gain a solid understanding of data center energy consumption and data centers energy indicators. Data centers energy indicators can be used to shape energy conservation strategies, and to determine the effectiveness of the measures to reduce energy consumption. In (Iulia and all) two important indicators that address energy efficiency are overviewed.

In a data center there are a lot of components that consume power (Krishna Kant). The visible components that consume power are: the racks, the servers, the network cables and the power cables, but what it is not visible are the CPUs running programs and the constant flow of information in and out.

Figure 1 outlines a typical data center energy consumption ratio (Neil Rasmussen). This is a power analysis of a typical highly available dual-power-path data center with N+1 computer room air conditioners (CRAC) units, operating at a typical load of 30% of design capacity.

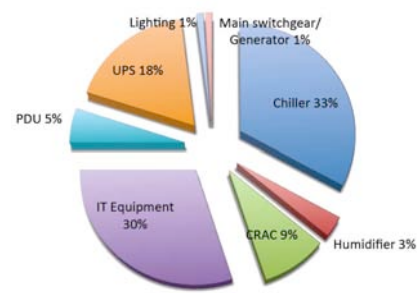

Fig.1. Power consumption in a typical data center [Neil Rasmussen]

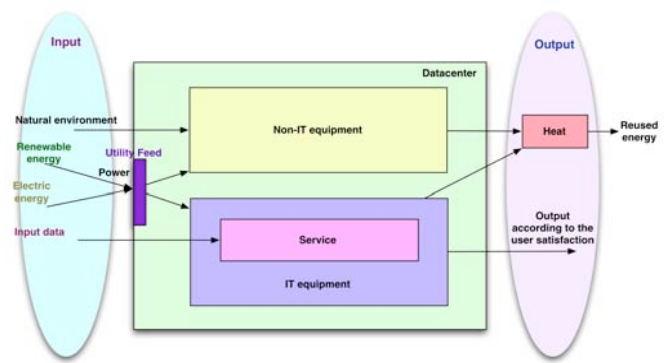

Fig. 2. A data center model [Green IT Promotion Council]

Using the analysis and the synthesis of informatics systems, which with the help of the technological process and its requirements for deployment can establish the IT architecture in terms of hardware and software (Sergiu Iliescu), a data center model can be obtained. A data center as shows in Figure 2 is divided into IT equipment such as servers and non-IT equipment such as air-conditioning, etc. Energy is required as input data into the data center and an output, that has to cope with the customer satisfaction, is obtained. The final energy is compound of the grid energy and renewable energy. Output is also accompanied by heat that can be reused. As output in the data center we can find also the energy losses. Since the natural environment (for example, a site location in a cold area) is a large factor of input, it was also incorporated into this model (Green IT Promotion Council).

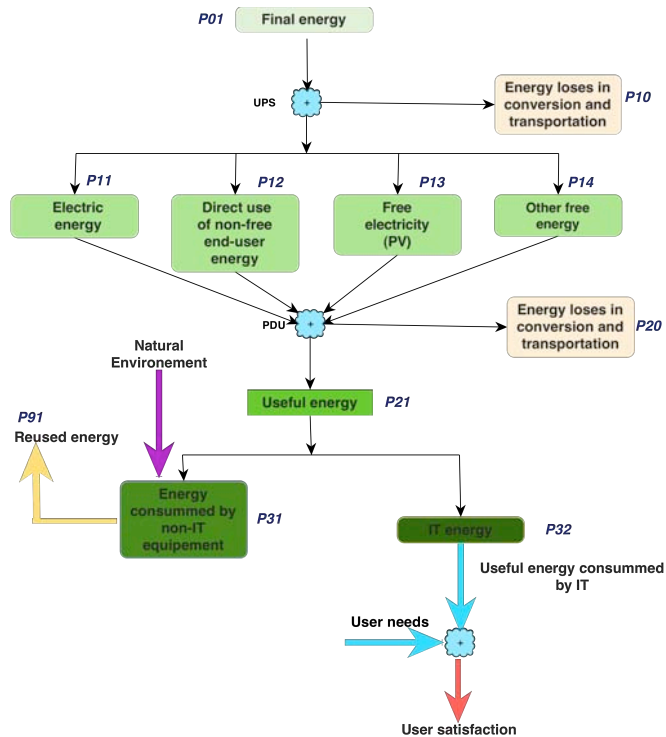

Fig 3. Energy repartition in a typical data center

In Figure 3, we propose an energy repartition in a typical data center. P01 represents the final energy (electricity) entering in the data center. The renewable energy sources (the so-called free energy) are also a final energy component. P10 are the losses in conversion and transport after the energy passes through the Power Utility and the uninterruptible power systems (UPS). Also, the remaining power can be divided into P11 the electric energy, P12 non-free end-user energy (like the cooling weather), P13 the free electricity (like photovoltaic energy, thermal energy) and P14 other free energy which can be captured by the heating, ventilation, and air conditioning (HVAC) systems. All this energy that passes through the power distribution unit (PDU) results in the P21- useful energy and also some energy losses. The useful energy can be divided into two parts: the energy used by the IT equipment and the energy used by the non-IT equipment. From the energy consumed by the non-IT equipment, a part can be re-used (P91). It can be reused, for example to warm up other building linked to the data center. We name useful energy consumed by the IT, the IT energy used to perform only the users required task. The Useful IT energy represents the energy that is used to compute only the tasks related with the user requests. The output of this has to satisfy the Service Level Agreement (SLA) (Edward Wustenhoff).

We can state that the two parties involved in the deployment of an application are the users and the data center mangers. Between the two of them an SLA is adopted. A SLA sets the expectations between the consumer and provider. It helps define the relationship between the two parties. It is the cornerstone of how the service provider sets and maintains commitments to the service consumer.

A good SLA addresses five key aspects (Edward Wustenhoff).

- What the provider is promising.
- How the provider will deliver on those promises.
- Who will measure delivery, and how.
- What happens if the provider fails to deliver as promised?
- How the SLA will change over time.

The SLA encapsulates many "behind-the-scenes" factors that contribute to energy consumption. Resuming, the SLA represents the user satisfaction.

The power consumption is not constant with time but varies according to different parameters. The main are the workload of the data center and the outside environment. Modeling the energy efficiency and the losses of the data centers equipment is a complex tasks and crucial assumptions yield great errors. First of all, the assumption that the losses associated to the power and cooling equipment are constant with time is wrong. It has been observed that the energy efficiency of this equipment is a function of the IT load and the non-IT equipment (devices responsible for cooling and power delivery) (Green IT Promotion Council).

## 2. THE DEGREES OF FREEDOM IN ENERGY EFFICIENCY

We call degrees of freedom, the number of independent pieces of information that can be found into an equation.

The facility benefits from degrees of freedom obtained from: the energy transport and conversion, the cooling and the heat evacuation and from the construction of the data center itself. On the other hand, many degrees of freedom can be insured by the usage of the IT. Below, we will speak about each degree of freedom.

In the quest for higher efficiency cooling system, it is imperative that the data center operates at the proper temperature. The pursuit of this goal should include considerations for the energy expended in the chilling process and the IT fan power, and it should also consider the best cooling solution and airflow management techniques to get there. One of the degree of freedom that was proposed in the literature is the increasing of the data center temperature. Some leading server manufacturers and data center efficiency experts share the opinion that data centers can run far hotter than they do today without sacrificing uptime, and with a huge savings in both cooling related costs and CO2 emissions. According to (Brian Renwick) it's estimated that data centers can save 4-5% in energy costs for every single degree increase in server inlet temperature. We think that supply air temperature should be raised but only after considering the implications to every piece of IT equipment. If a data center is successful in raising the delivers temperatures, there are large saving in the energy at the chiller, potential savings in moisture control, and there is a potential increase in the number of hours that economizer modes can be used (Davis Moss and all).

Others papers (Victor K. Lee) show how to place equipment in the data center in order to be as effective as possible in releasing the heat. The manner in which the data center was built affects the data center energy efficiency. The goal of all the researches done in this area is to find the best data center structure that ensures an efficient release of the heat from the IT and electrical equipment. In conclusion, the development of a smoother cooling controller in addition with smart equipment positioning in order to minimize cooling consumption can increase and sustain the process of optimization and management of the energy into a data center.

The transport energy doesn't have dynamic degrees of freedom because; the reliability and the solution to optimize the consumption of those equipments cannot be influenced by the data center designer/operator. The energy optimization is fulfilled with an optimal sizing and configuration, which can guarantee an efficient relation between the different data center infrastructures (tiers).

If we consider the IT Efficiency, we can state that one promising approach to increase energy efficiency could be virtualization and server consolidation. Consolidation and virtualization (Vinicius Petrucci and all) introduce additional degrees of freedom that are not available under traditional operating models.

Furthermore, another degree of freedom can be obtained if server consolidation is combined with dynamic voltage and frequency scaling capabilities offered by current processors to get even better results. Remarkable work has also been made from the point of view of storage data; we include here the variation of disc speeds to allow gain in energy under stress and rapid recovery in terms of performance.

### 3. DYNAMIC PRINCIPLES OF ENERGY MANAGEMENT

Server consolidation is based on the observation that many enterprise servers do not maximally utilize the available server resources all of the time. Co-locating applications as is showed in (Akshat Verma and all, 2009), perhaps each service in individual virtual machines, allows for a reduction in the total number of physical servers as well as the increasing of energy efficiency. In general, to allow hosting multiple independent applications, these platforms rely on virtualization techniques to enable the usage of different virtual machines (i.e., operating system plus software applications) on a single physical server. Virtualization provides a means for server consolidation and allows for on demand migration and allocation of these virtual machines, which run the applications, to physical servers.

Applications run in virtualized environments allowing the dynamic consolidation of workloads. Under these circumstances there will be a designated nucleus of highly utilized servers actually running the workloads, very efficiently due to the high loading. The other servers are basically idle and can be put into a low energy, sleeping state until needed.

Intelligently turning of spare servers that are not being used is an obvious way to reduce both power and cooling costs while maintaining good performance levels (Josep Ll. Berral and all). This approach solves some interesting challenges; less hardware is required, less electrical consumption is needed for server power and cooling and less physical space is required. In conclusion, an energetic gain will be obtained because of the possibility to turn on and off the virtual machines.

In recent literature three types of consolidation can be found: static, semi-static and dynamic consolidation. In static consolidation, applications (or virtual machines) are placed on physical servers for a long time period (e.g. months, years), and not migrated continuously in reaction to load changes. Semi-static refers to the mode of consolidating these applications on a daily or weekly basis. On the other hand, dynamic consolidation requires a runtime placement (a couples of hours) manager to migrate virtual machines automatically in response to workload variations. One of the purposes of virtualization and consolidation is to reduce the power consumption of a data center while respecting the different SLA-s.

In the process of consolidation of several tasks, distributed among a set of machines, into as few machines as possible without degrading excessively the execution of these jobs, several scheduling policies could be applied. The scheduling policies are usually used to balance the systems load effectively or achieve a target quality of service. The need for a scheduling algorithm arises from the requirement for most modern systems to perform multitasking (execute more than one process at a time) and multiplexing (transmit multiple flows simultaneously). In (Josep Ll. Berral and all) several scheduling policies are presented that can be used in order to assign new jobs in the system to available machines and redistribute jobs being executed in order to make some machines idle and then turning them off. Notice that turning machines on again is not a free and instantaneous process and this overhead, which can take more than a minute, must be taken into account.

Below we present several traditional scheduling policies, that can be applied to manage data or task inside one single non-virtualized server or one virtual machine including: round robin, backfill and dynamic backfill. The first two techniques are static scheduling policies and the last one represents a dynamic policy that uses task migration.

Round-robin (RR) is one of the simplest scheduling algorithms for processes in an operating system, which assigns time slices to each process in equal portions and in circular order, handling all processes without priority (also known as cyclic executive). Round robin scheduling is both simple and easy to implement, and starvation-free.

The main goal of backfill algorithm is to try and fit small jobs into scheduling gaps. It improvises the system utilization by running low priority jobs in between the high priority jobs. Runtime estimate of small jobs provided by the users are required by the scheduler in order to use backfill.

Dynamic Backfilling that is able to move (i.e. migrate) tasks between nodes in order to provide a higher consolidation level. When tasks enter or exit the system, it checks if any tasks should be moved to other nodes according to different parameters such as the system occupation, current job performance, or expected user SLA satisfaction. While Dynamic Backfilling performs well when having precise information (as shown in the evaluation), other policies are necessary when information is incomplete or imprecise. Virtualization has initially been introduced to allow better development of technical capacity, to reduce the cost but also the total operating space. Over time virtualization and its various developments including the possibility of migration, of one virtual machine from one host to another, have opened a new research area in terms of energy optimization.

The major problem that dynamic consolidation/ VM migration has to deal with is to decide what the best location for executing a new job is, depending on the resources it requires in order to fulfill its SLA and according with the providers interests on power efficiency, reliability, etc.

There are several algorithms (Akshat Verma, 2008) that propose an energy management method for migration tasks. A good algorithm it should take into consideration a number of requirements or constraints as (Victor K. Lee):

- Hardware and software requirements: for each VM, it should be checked if the host is able to hold it, by evaluating the hardware (it has the required system architecture, the required type and number of CPUs, etc.), the software (it has the capability to execute a given software, it uses a given hypervisor, e.g. XEN, KVM, . . .).
- SLA requirements: it has to be ensured that the VM can get the amount of resources it requires to fulfill the SLA.
- Time requirements: the algorithm should take into consideration the virtualization overheads: the creation overheads (which is the time required to create and start a VM before it is ready to run tasks) and the migration overheads (which is the time incurred when moving a running VM between two different nodes.

The mechanism most frequently encountered in the dynamic consolidation techniques is a mechanism (Victor K. Lee) that uses a matrix. The matrix coefficients are used to weight the various possible choices and follow the optimal one. The scheduling policies are usually used to balance the systems load effectively or achieve a target quality of service.

The need for a scheduling algorithm arises from the requirement for most modern systems to perform multitasking (execute more than one process at a time) and multiplexing (transmit multiple flows simultaneously).

An interesting optimization approach can be persuade if instead of resizing the virtual machines for the worst case scenario, a method used on large scale in the literature, we would dynamically adjust the virtual machines allocation to physical servers with different reliabilities. The major problem that dynamic migration has to deal with it to decides what the best location for executing a new job is, depending on the resources it requires in order to fulfill its SLA and according with the providers interests on power efficiency, reliability, etc. A good algorithm it should take into consideration a number of requirements or constraints as shown in (Vinicius Petrucci and all). The mechanism most frequently encountered in the dynamic consolidation techniques is a mechanism (Dara Kusic) that uses a matrix. Matrix coefficients are used to weight the various possible choices and follow the optimal one.

In the often-encountered approaches we find the notion of "online optimization". Online optimization means that the optimization calculations are done at the same time as the flow is changing. This approach has the merit of being robust to uncertainties, but if we view this approach from a global perspective over a horizon of time we can state that these operations improvements are not optimal. The choices are optimal only at that specific moment of time and with the specific constraints, but if we want to have an optimal decision over a long horizon of time we need to predict the virtual machine's loads. If we use the load predictions we can have better performance in term of optimization and we can also correct the uncertainties prediction in an online manner. This approach is already implemented in another areas like energy efficiency in smart houses.

Furthermore, another degree of freedom can be obtained if server consolidation is combined with dynamic voltage and frequency scaling capabilities offered by current processors to get even better results. Remarkable work has also been made from the point of view of storage data; we include here the variation of speed discs to allow a gain in energy under stress and rapid recovery in terms of performance (Vinicius Petrucci and all).

We further think, that another degree of freedom can be introduced if the user satisfaction related with the Service Level Agreement is time varying. Future contribution will take into account new degree of freedom, which are not exploited until today as: the fact that some tasks can be moved in time and that the SLA can be dynamic for specific applications. In addition to this, we can consider that there are multiple tasks that can be repetitive and therefore predicted. From an other point of view we seek, also to minimize the energy loses due to the sizing of virtual machine within the worst case (peak) of workload.

Another good optimization idea, if we can provide a varying satisfaction related with the SLA, is the shifting of the IT workloads to ensure that they consume power during the time period where CO2 emission is lower.

4. CONCLUSIONS

In this paper we presented the electric flow in a data center as and the different degrees of freedom encountered into a data center. We have described the process of energy management as well as different principals of energy management.

Energy optimization of data centers represents a major concern and for that several methods have been developed and implemented. We think that a dynamic predictive control and allocation can represent the solution for increasing energy efficiency. Monitoring the proper inputs of the data center model, the creation of a new instance problem over time, and the anticipative management using prediction would be an important source of efficiency gain.

Concluding, our future work will deal with dynamical and anticipative energy management of data centers taking into consideration the satisfaction related to a variable Service Level Agreement. We strongly feel that an anticipative management using prediction will avoid the resizing of the data center for the worst-case workload.

REFERENCES

Akshat Verma (2008), Mapper: power and migration cost aware application placement in virtualized systems, *Middleware08, pages 243-264*.

Akshat Verma, Gargi Dasgupta, Tapan Kumar Nayak, Pradipta De, Ravi Kothari (2009), Server Workload Analysis for Power Minimization using Consolidation, *USENIX'09 Proceedings of the Conference*

Brian Renwick (2009), Surprising Cost Savings Through Server Room Temperature Monitor Strategies, *The American Society of Heating, Refrigerating and Air-Conditioning Engineers (ASHRAE)*

Davis Moss, John H. Bean (2009), Energy Impact of Increased Server Inlet Temperature, *APC by Schneider Electric - White Paper 138*

Dara Kusic (2009), Power and performance management of virtualized computing environments via lookahead control, Cluster Computing, *Journal Cluster Computing, Volume 12 Issue 1*

Edward Wustenhoff (2002), Service Level Agreement in Data Center, *Sun BluePrints OnLine*

Green IT Promotion Council (2010), Japan, Concept of New Metrics for Data Center Energy Efficiency Introduction of Data Center Performance per Energy - [DPPE], *Green IT Promotion Council*

Iulia Dumitru, Stéphane Ploix, Ioana Făgărăşan, Sergiu Iliescu (2010), Data Center Control: Guidelines for Obtaining Energy Efficiency, 18th INTERNATIONAL CONFERENCE ON CONTROL SYSTEMS AND COMPUTER SCIENCE, Bucharest, Romania.

Josep Ll. Berral, Inigo Goiri, Ramon Nou, Ferran Juli, Jordi Guitart, Ricard Gavald and Jordi Torres (2010), Towards energy-aware scheduling in data centers using machine learning, Energy efficient data center technology, *e-Energy '10 Proceedings of the 1st International Conference on Energy-Efficient Computing and Networking*

Krishna Kant (2009), Data center evolution: A tutorial on state of the art, issues, and challenges, *Computer Networks 53*, 29392965

Neil Rasmussen (2009), Electrical Efficiency Modeling for Data Centers, *American Power Conversion White Paper 113*

Ricardo Bianchiniy, Ram Rajamony (2004), Power and Energy Management for Server Systems, *IEEE Computer, Volume 37, Number 11. Special issue on Internet data centers*

Sergiu Stelian Iliescu, Ioana Făgărăşan, Dan Pupăză (2006), Analiza de sistem în informatica industrială, *Editura AGIR*, Bucureşti, ISBN(10): 973-720-091-8

Vinicius Petrucci, Orlando Loques, Daniel Mosse (2010), A dynamic Optimization model for power and performance management of virtualized clusters, *e-Energy '10 Proceedings of the 1st International Conference on Energy-Efficient Computing and Networking*

Victor K. Lee (2010), Power and Cooling Architecture in Future Data Center, *The Applied Power Electronics Conference and Exposition*